# Estimation of cold plasma outflow during geomagnetic storms

S. Haaland[1,2], A. Eriksson[3], M. André[3], L. Maes[4], L. Baddeley[5], A. Barakat[6], R. Chappell[7], V. Eccles[6], C. Johnsen[8], B. Lybekk[9], K. Li[2], A. Pedersen[9], R. Schunk[6], and D. Welling[1,10]

[1]Birkeland Centre for Space Science, University of Bergen, Bergen, Norway, [2]Max-Planck Institute for Solar Systems Research, Göttingen, Germany, [3]Swedish Institute of Space Physics, Uppsala, Sweden, [4]Belgian Institute of Aeronomy, Brussels, Belgium, [5]Department of Arctic Geophysics, University Centre in Svalbard, Longyearbyen, Norway, [6]Center for Atmospheric and Space Sciences, Utah State University, Logan, Utah, USA, [7]Science and Research Communications, Vanderbilt University, Nashville, Tennessee, USA, [8]Department of Geophysics, University of Oslo, Oslo, Norway, [9]Department of Physics, University of Oslo, Oslo, Norway, [10]Department of Atmospheric, Oceanic and Space Sciences, University of Michigan, Ann Arbor, Michigan, USA

**Abstract** Low-energy ions of ionospheric origin constitute a significant contributor to the magnetospheric plasma population. Measuring cold ions is difficult though. Observations have to be done at sufficiently high altitudes and typically in regions of space where spacecraft attain a positive charge due to solar illumination. Cold ions are therefore shielded from the satellite particle detectors. Furthermore, spacecraft can only cover key regions of ion outflow during segments of their orbit, so additional complications arise if continuous longtime observations, such as during a geomagnetic storm, are needed. In this paper we suggest a new approach, based on a combination of synoptic observations and a novel technique to estimate the flux and total outflow during the various phases of geomagnetic storms. Our results indicate large variations in both outflow rates and transport throughout the storm. Prior to the storm main phase, outflow rates are moderate, and the cold ions are mainly emanating from moderately sized polar cap regions. Throughout the main phase of the storm, outflow rates increase and the polar cap source regions expand. Furthermore, faster transport, resulting from enhanced convection, leads to a much larger supply of cold ions to the near-Earth region during geomagnetic storms.

## 1. Introduction

Ions of ionospheric origin are believed to be a significant contributor to the magnetospheric plasma population [*Shelley et al.*, 1982; *Horwitz*, 1982; *Chappell et al.*, 1987, 2000; *André and Cully*, 2012]. Major ionospheric outflow regions include the auroral zone [e.g., *Wahlund and Opgenoorth*, 1989; *Winser et al.*, 1989; *Yau et al.*, 1993; *André et al.*, 1998; *Wilson et al.*, 2001], the cusp [e.g., *Yau et al.*, 1985b; *Lockwood et al.*, 1985b, 1985a; *Yau and Andre*, 1997], and the polar cap area [e.g., *Brinton et al.*, 1971; *Chandler et al.*, 1991; *Abe et al.*, 1993; *Moore et al.*, 1997; *Su et al.*, 1998].

Above the open polar cap regions, where no hydrostatic equilibrium can be established, low-energy photoelectrons can escape the Earth's gravitational field. Consequently, a spatial separation between the light electrons and the heavier ions (mainly $H^+$, $He^+$ and $O^+$) arises, and an ambipolar electric field which eventually accelerates lighter ions upward is set up. Once free of the gravitational potential, the polar wind expands at supersonic speed along the magnetic field into the magnetotail lobes. This outflow of plasma from the polar cap area is often referred to as the polar wind [*Banks and Holzer*, 1968; *Axford*, 1968; *Yau et al.*, 2007].

The ambient electric field associated with the polar wind is very small. Simulations by *Su* [1998] suggest that the total potential drop of a few tens of Volts extends over an altitude of several Earth radii. Consequently, direct observations of this potential drop is not possible, and only indirect methods can be used. *Kitamura et al.* [2012] used the photoelectron flow data from the Fast Auroral SnapshoT satellite during geomagnetically quiet periods and inferred that potential drops above the satellite (approximately 3800 km altitude) were typically around 20 V. Inferred potentials below the satellite were much lower, only 1–3 V.

In a later follow-up study [*Kitamura et al.*, 2013], using data obtained during the main and early recovery phases of geomagnetic storms, they reported typical potential drops of 5 V or less—i.e., much smaller than during quiet periods. They attributed this to stronger convection of ions from the cusp area during disturbed conditions, which will effectively reduce the effect of the photoelectrons.

Such low potential drops mean that little energy is available to accelerate ions in this region. Additional acceleration, like, e.g., centrifugal acceleration [e.g., *Cladis*, 1986; *Nilsson et al.*, 2008, 2010], is not very effective over short distances [*Demars et al.*, 1996]. Unlike the cusp and cleft regions, there is no significant energization from solar wind driven Poynting flux [e.g., *Zheng et al.*, 2005; *Strangeway et al.*, 2012] or wave activity [e.g., *Wahlund et al.*, 1992; *Bouhram et al.*, 2004]. Thus, ions emanating from the polar cap will not gain significant energy as they travel outward — they will remain cold.

It is notoriously difficult to conduct in situ measurements of the cold part of the outflowing ion population. In the tenuous plasma regions of the Earth's magnetosphere, the spacecraft voltage often reach several tens of volts positive due to photoelectron emissions from the spacecraft surface. This spacecraft potential will prevent low-energy ions from reaching spacecraft sensors. Unless the effects of spacecraft charging can be eliminated, cold ions therefore remain invisible for particle detectors.

Attempts to bypass this problem typically involve some form of active spacecraft potential control. *Su et al.* [1998] used particle measurements from the Polar spacecraft and presented a study of cold ion outflow during a limited time period when the on board Plasma Source Instrument was operating and kept the spacecraft potential at a few volts. They were then able to observe and characterize the polar wind outflow at high altitudes for this time period. The Cluster spacecraft [*Escoubet et al.*, 1997] which forms the observational basis for the present study also has an Active Spacecraft Control instrument [see *Riedler et al.*, 1997], but to our knowledge no specific study focusing on polar wind or ion outflow has systematically utilized this.

*Engwall et al.* [2006] presented a completely different approach to cold ion outflow detection. By utilizing data from two independent electric field instruments, they were able to exploit the spacecraft charging and measurements from the two experiments to derive densities and outflow velocities of cold plasma. This technique has also been applied by a number of follow-up studies, e.g., *Engwall et al.* [2009a], *Haaland et al.* [2012a, 2012b], *Li et al.* [2012, 2013], *André et al.* [2015], and S. Haaland et al. (Magnetosphere-ionosphere coupling in the solar system, chap. 1, in review, 2015) and will also be applied in the present study. The principles of this method will be described in section 2 of the present paper.

The motivation for this paper was a call from the Geomagnetic Environment Modelling (GEM) core group to provide observational inputs for benchmarking, parametrization, and verification of geophysical models valid during geomagnetic storms. During the years 2013–2015, the core group set up a project in which they selected three events to study closely: an idealized synthetic event and two real geomagnetic storm events. Numerical modelers were invited to simulate each and compare their results to other models. Additionally, data experts and experimentalists were invited to share observations of the real-world events and contribute to data-model comparisons. The present paper reports on observations of cold ion outflow which may be useful for this purpose.

The paper is organized as follows: In section 2, we explain why cold ion measurement are difficult and how the instrumentation on board Cluster is used to bypass the spacecraft charging problem. We also provide a description of the data set used for this study and its characteristics. Section 3 presents the results, and section 4 discusses the implications. Finally, section 5 summarizes the results.

## 2. The Cold Ion Detection Challenge

A spacecraft traversing the Earth's high-altitude polar cap and magnetically connected lobe region will be exposed to solar illumination. This illumination, in particular, in the extreme ultraviolet (UV) range, will cause photoionization of the spacecraft surface area. In the tenuous plasma of the polar cap and lobes, the photoelectrons cannot easily be replenished. Consequently, the spacecraft will be positively charged (see details in, e.g., *Pedersen et al.* [2001, 2008] and *Lybekk et al.* [2012]). Unless this charging can be prevented, this will cause problems for low-energy plasma measurements.

For Earth, escape energies for protons and oxygen are around 0.6 and 10 eV, respectively. Typically, there are no strong acceleration mechanisms above the polar cap region, and a substantial amount of these cold outflowing ions will remain "cold" as they move outward. If the energy of these ions is below the spacecraft potential energy ($eV_{SC}$, where $e$ is elementary charge and $V_{SC}$ is the spacecraft potential relative to the ambient plasma), these ions will not be able to reach the detectors and are thus "invisible" as illustrated in Figure 1.

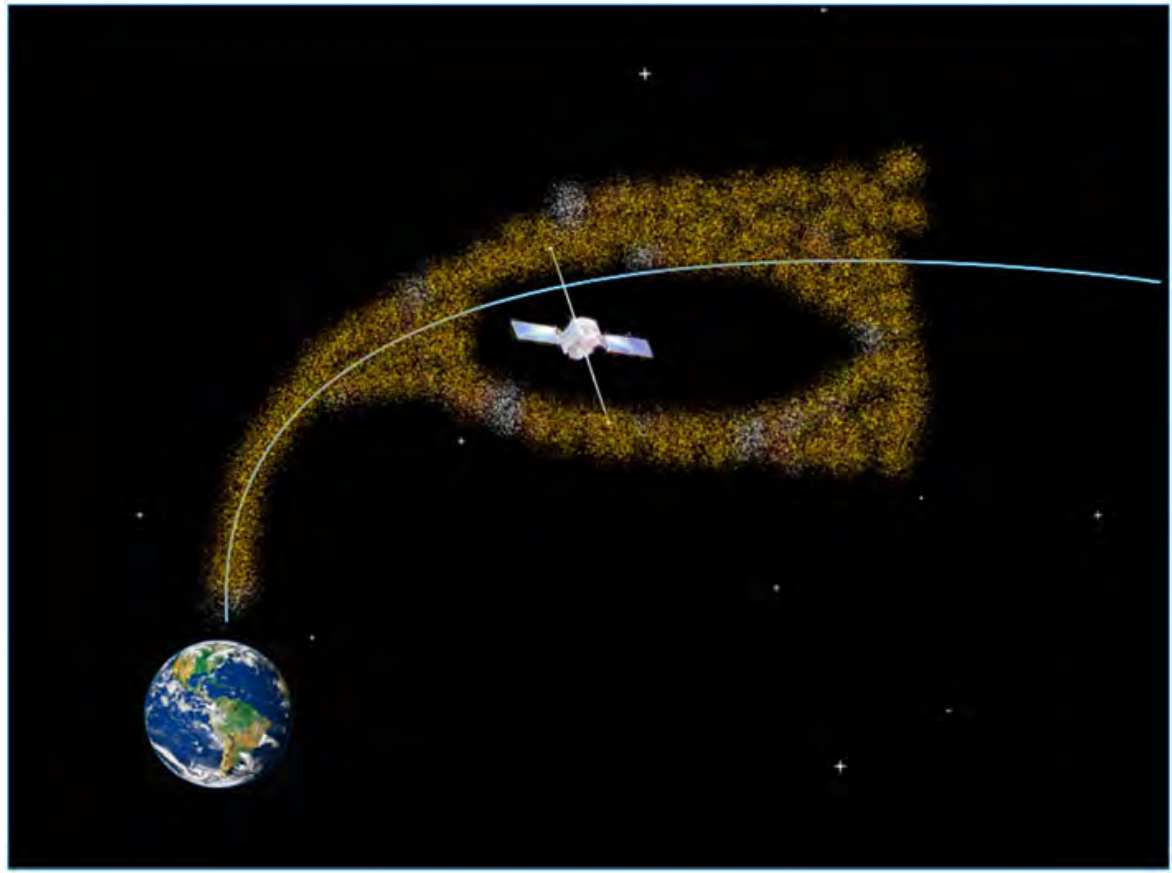

**Figure 1.** Illustration of shielding due to spacecraft charging. Low-energy ions emanating from the polar cap region travel upward along the magnetic field lines. Due to positive spacecraft charging, ions with energies below the spacecraft potential energy will not reach particle detectors on board the satellite—they remain invisible. A wake will be formed downstream of the spacecraft.

Remote sensing of ion outflow is also difficult. Ground-based measurements, e.g., incoherent scatter radars can only measure up to about 1000 km altitude. Vertical upward motion, at these altitudes termed upwelling, often goes along with downward vertical motion. It is thus difficult to assess how much plasma actually reaches escape velocity and actually escapes the Earth's gravitational field. Low-orbit satellites, although less affected by spacecraft charging, have similar issues.

### 2.1. Utilizing Spacecraft Potential and Wake

The present study is based on observations from the Cluster constellation of spacecraft [*Escoubet et al.*, 1997]. A unique feature of the Cluster mission is the combination of two complementary electric field experiments, the Electric Field and Wave experiment (EFW) [see *Gustafsson et al.*, 2001] and the Electron Drift Instrument (EDI) [see *Paschmann et al.*, 1997; *Quinn et al.*, 2001] This combination is the key element for the technique to estimate cold ion flux developed by *Engwall et al.* [2006].

EFW is a classic double probe instrument, consisting of two pairs of boom-mounted spherical probes. The probe to probe distance is approximately 88 m for each pair. This arrangement provides measurements of the electric field in the satellite spin plane. Assuming no or negligible electric potential drop along the magnetic field ($E_\perp \gg E_\parallel$), the full 3-D electric field can be estimated provided that the spin axis is not parallell to the magnetic field.

EDI is based on the drift of an electron gyrocenter in the presence of external forces. Each Cluster spacecraft is equipped with two EDI gun/detector units, each emitting a modulated electron beam with a fixed energy. (The beam energy can be switched between 500 eV and 1 keV to measure the effect of magnetic gradients, but as these are usually small compared to the local electron gyroradius, and the beam energy is typically kept fixed at 1 keV). The firing direction of this beam is continuously controlled through a servo loop so that the coded beam returns to the detector unit. Gyrocenter position and motion can then be determined from triangulation (or, in some regions, from the time of flight of the emitted electrons). For a known magnetic field with negligible gradients, the gyrocenter drift of the emitted beam is proportional to the convective electric field. In regions with fairly stable magnetic field, and low-electron background plasma, EDI provides the full 3-D convective electric field with very high accuracy. Strong variations in the magnetic fields or strong gradients can prevent successful tracking and no valid $E$ field can be calculated. Likewise, a strong electron background density can lead to an attenuation of the modulated beam, and tracking is lost.

Cluster consists of four identical spacecraft flying in a formation with variable separation distance. In the community, the spacecraft are simply referred to as C1, C2, C3, and C4. All four spacecraft are equipped with identical instruments, but EDI is only fully operational on C1 and C3.

### 2.1.1. Cold Plasma Density
With knowledge of surface properties and surface area and a known solar illumination, it is possible to use the spacecraft potential to estimate the ambient electron density, and thus the plasma density [e.g., *Pedersen et al.*, 2001; *Lybekk et al.*, 2012, and references therein]. In general, a relation of the form

$$N_e = Ae^{-BV_{SC}} + Ce^{-DV_{SC}} \qquad (1)$$

exists, where $N_e$ is the sought after electron density, $V_{SC}$ is the spacecraft potential relative to the ambient plasma. The coefficients $A$, $B$, $C$, and $D$ are determined from calibrations against other measurements and implicitly contain information about solar illumination and spacecraft surface properties.

### 2.1.2. Cold Ion Bulk Velocity
If the bulk energy, $E_K$, of the cold ions flowing across the spacecraft is larger than their thermal energy, $kT_i$, i.e., the following inequality exists

$$kT_i < E_K < eV_{SC}, \qquad (2)$$

a wake void of ions will be formed downstream of the spacecraft. Electrons, however, with their higher mobility (typically $kT_e >> E_{Ke}$), will be able to fill the wake. Consequently, an electric field, $\vec{E}^W$ along the bulk flow direction, $\vec{u}$ will arise:

$$\vec{E}^W = g\vec{u}, \qquad (3)$$

where the scaling factor, $g$, is a function of the local plasma parameters and can be experimentally determined [*Engwall et al.*, 2006].

The size of the wake is comparable to the boom-to-boom scale size of the spacecraft but much smaller than the gyroradius of the 1 keV electron beam emitted by EDI, which is of the order of several kilometers for the typical magnetic field strength in the lobes. The probe-based measurements from EFW will thus be influenced by this artificial electric field, whereas EDI is not affected. The wake electric field can then be expressed as a deviation between the wake influenced electric field measured by EFW, $\vec{E}^{EFW}$, and the real, unperturbed ambient electric field $\vec{E}^{EDI}$:

$$\vec{E}^W = \vec{E}^{EFW} - \vec{E}^{EDI} = g\vec{u}. \qquad (4)$$

Note that the perpendicular part of the bulk flow, $\vec{u}_\perp$, is obtained directly from the EDI measurements $\vec{u}_\perp = \vec{E}^{EDI} \times \vec{B}/B^2$. The parallel component of $u$ can then be obtained by decomposition $\vec{E}^W$ into two spin-plane component, $E_x^W$ and $E_y^W$. An explicit expression for the parallel bulk velocity of the cold ions can then be obtained

$$u_\parallel = \frac{E_x^W u_{\perp,y} - E_y^W u_{\perp,x}}{E_y^W B_x - E_x^W B_y}\vec{B}, \qquad (5)$$

where $B$ is the magnetic field.

Note that wake formation as such is not exclusive to the polar cap or lobe regions [e.g., *Whipple et al.*, 1974, and references therein], but the combination of the two electric field measurements on board Cluster has made determination of the bulk velocity possible for the first time.

### 2.1.3. Cold Ion Outflow Flux
From the above equations (1) and (5), the flux of cold ions at the spacecraft position can now be determined

$$f_\parallel = N_e * u_\parallel. \qquad (6)$$

Using flux conservation consideration and flux tube cross section from a magnetic field model, we can now scale this flux to ionospheric altitudes. Particle tracing can be used to determine the source region or fate of the outflowing ions [e.g., *Cully et al.*, 2003; *Li et al.*, 2012, 2013].

### 2.2. Limitations of the Wake Method

From the above derivation, one notes that it is not possible to distinguish between different ion species. Nor is any distinction between different ion charge states possible, so singly ionized ions are assumed in the above derivations. The wake method is more sensitive to lighter ions, as these are more affected by the wake. Observations by *Su et al.* [1998] indicate that hydrogen is the dominant species in low-energy outflow from the polar cap region. Nevertheless, in *Engwall et al.* [2009a] and *André et al.* [2015], the derived densities have been lowered by a factor of 0.8 to account for the presence of heavy ions. In reality, the abundance of heavier ions, typically oxygen, in the outflow varies both with geomagnetic activity and source location. Oxygen is more likely to emanate from the cusp and auroral zone [e.g., *Yau and Andre*, 1997; *Lockwood et al.*, 1985b, 1985a].

The inequality in equation (2) limits the temperature and bulk energy ranges of the ions possible to detect. Also, since the velocity determination rests on the identification and characterization of a downstream wake (which is not always observed—even in the polar cap and lobe regions), the data set is not continuous in time but consists of individual intermittent records. Furthermore, the bulk flow direction should have a significant component along the spin plane of the spacecraft. Otherwise, the EFW probes will not be able to measure the wake field. This is usually no issue in the lobes, where the magnetic field is stretched out, but can be an issue closer to Earth.

As with any collection of experimental data, there are uncertainties related to measurements, methodology, and the underlying assumptions. *Engwall et al.* [2009a] estimated that error due to methodology is of the order of ± 40% or less for velocity calculations and of the order of 20% for electron density calculations.

### 2.3. Source of Cold Ions

In order to calculate the total outflow of cold ions, we also need to know the area of the source region, i.e., essentially the area of open magnetic flux in both hemispheres. In their initial estimate of outflow rates, *Engwall et al.* [2009a] used a fixed polar cap boundary located at 70° magnetic latitude. Neither expansion and contraction of the polar cap nor any spatial inhomogeneities were taken into account. Later, *Haaland et al.* [2012a] used a variable polar cap area, parametrized by the solar wind input energy after a method developed by *Sotirelis et al.* [1998]. They noted large variations in the source area due to the expansion and contraction of the polar cap in response to geoactivity.

*Li et al.* [2012] performed particle tracing to generate maps of the source area and could thus also address any inhomogeneities in the source. Their results confirmed that the open polar cap is the primary source of the cold ions, but they also found enhanced outflow from a region near the cusp and a region near the nightside auroral zone during disturbed conditions. Other than that, no significant day-night asymmetry in the outflow was observed. Around equinox, most of the polar cap ionosphere is illuminated at least parts of the day, both in Northern and Southern Hemisphere, and this may explain the lack of a pronounced dayside-nightside difference in outflow. Another factor is that convection and vertical winds will cause some redistribution of the cold ions between the peak ionization layers (the *D*, *E*, and *F* layers of the ionosphere) and the topside ionosphere. There is probably also mixing of ions from different ionospheric regions along the transport path to the lobes where they are detected by Cluster.

Since the purpose of the present paper is to address the cold outflow during geomagnetic storms, i.e., limited time periods, we use subsets of the full data set. The method of *Li et al.* [2012] is therefore not applicable, since it requires full spatial coverage in order to determine the size of the source area. We therefore use the procedure outlined in *Milan* [2009], which provide a proxy for the open flux area as function of dayside reconnection electric field, $\Phi_D$, and the *Dst* index (*Milan* [2009] actually uses the *SYM-H* index: see section 2.4 for a discussion of various indices to characterize geomagnetic storms). The dayside reconnection field is a measure of opening of flux at the dayside magnetopause, and the *Dst* index provides a similar proxy for flux closure on the nightside. Any imbalance between these two processes will lead to an expansion or contraction of the area of open flux.

Based on 40,000 independent observations of auroral images, *Milan* [2009] came up with the following relation between $\Phi_D$, *Dst*, and the auroral oval radius:

$$\lambda = 18.2 - 0.038 \, Dst + 0.042 \, \Phi_D, \tag{7}$$

where $\lambda$ is the radius (in degrees) of the auroral oval. Note that the auroral oval is not necessarily centered around any geomagnetic axis, so $\lambda$ is, in general, not the colatitude of the oval location. The dayside reconnection electric field is given by

$$\Phi_D = 2.75\, R_E\, V_{SW} \sqrt{B_Y^2 + B_Z^2}\, \sin^2(\theta/2), \qquad (8)$$

where $2.75\, R_E$ is a characteristic length scale, $V_{SW}$ is the solar wind bulk flow speed, $B_Y$ and $B_Z$ are components of the interplanetary magnetic field, and $\theta$ is the IMF clock angle, here defined as $\theta = \mathrm{acos}\left(B_Z/\sqrt{B_Y^2 + B_Z^2}\right)$.

Expression (7) refers to the peak of the auroral oval as identified from the images. Our source area is poleward of the auroral oval. We have assumed an average auroral oval width of approximately 5° in latitude, and therefore shifted the open-closed boundary poleward by using a 2.5° smaller radius for our source area. This poleward shift is consistent with open-closed boundaries determined from in situ particle measurements (see, e.g., discussion in *Boakes et al.* [2008], and references therein.)

At 1000 km altitude, 1° in latitude corresponds to approximately 128 km. The size of the source area in one hemisphere, $A$, can thus be expressed as

$$\begin{aligned}A &= \pi\, [\lambda \ast 128\,\mathrm{km}]^2 \\ &= \pi\, [(15.7 - 0.038\, Dst + 0.042\, \Phi_D) \ast 128\,\mathrm{km}]^2. \end{aligned} \qquad (9)$$

All of the above used quantities are known and available from our data set, and we can thus calculate the total area (both hemispheres) of the source of cold ions for a given combination of disturbance level and solar wind input.

### 2.4. Data Set Overview

In the present study, we combine several data sets, already described in previous publications.

First, we use an extended set of wake observations which provides parallel bulk flow velocities of the cold ions. This extended data set contains observations from Cluster C1 and C3. The data set was prepared and presented by *André et al.* [2015] and is based on an earlier, similar data set prepared by *Engwall et al.* [2009a]. The new wake data set contains about twice as many observations as the earlier set of observations (approximately 350,000 records in *André et al.* [2015] versus 176,000 records in *Engwall et al.* [2009a]). In addition to the full processing for Cluster spacecraft 1, the new data set contains five more years of observations, and thus cover almost a full solar cycle (2001–2010). Since the data set relies on the detection and characterization of an electrostatic wake, observations are only available intermittently, and only when Cluster traverses the high-latitude nightside lobes, i.e., in the period July to November each year. Several studies have utilized these wake data sets, e.g., *Engwall et al.* [2009b], *Nilsson et al.* [2010, 2012], *Haaland et al.* [2012a, 2012b], S. Haaland et al. (in review, 2015), and *Li et al.* [2012, 2013].

Cold ion densities are based on measurements reported by *Lybekk et al.* [2012]. This data set contains data from all four Cluster spacecraft for the period 2001–2010. In the present study, we only use data from Cluster C1 and C3 since these are the only spacecraft with usable wake observations. Convection measurements, used to study the transport of plasma, are discussed in *Haaland et al.* [2008, 2009] and based on EDI measurements from Cluster C1 and C3 for 2001–2010.

In addition to the above, we also use auxiliary parameters such as solar wind data, geomagnetic indices, and the $F_{10.7}$ index — a daily proxy for solar UV illumination.

To characterize the geomagnetic disturbance level, we use SuperMag-based partial ring current indices (*SMR*) [see *Gjerloev*, 2012; *Newell and Gjerloev*, 2012]. The standard *SMR* index is essentially the same as the Disturbed Storm Time index (*Dst*) [see *Sugiura*, 1964] or the similar *SYM-H* index [*Wanliss and Showalter*, 2006] but is constructed from a larger number of observatories. *SYM-H* and *SMR* are available at 1 min time resolution, whereas the original *Dst* index is an hourly average. All three indices are measurements of perturbations in the horizontal component of the Earth's magnetic field around equatorial latitudes and provide a proxy for the energy in the Earth's ring current.

Density and convection measurements are available at almost all times when Cluster is in the lobes or polar cap regions. However, a full characteristics of the cold ion outflow is only possible when both wake observations and convection measurements are present. Figure 2 shows the *SMR* index for 2001–2010, with periods where wake measurements could be utilized indicated in red.

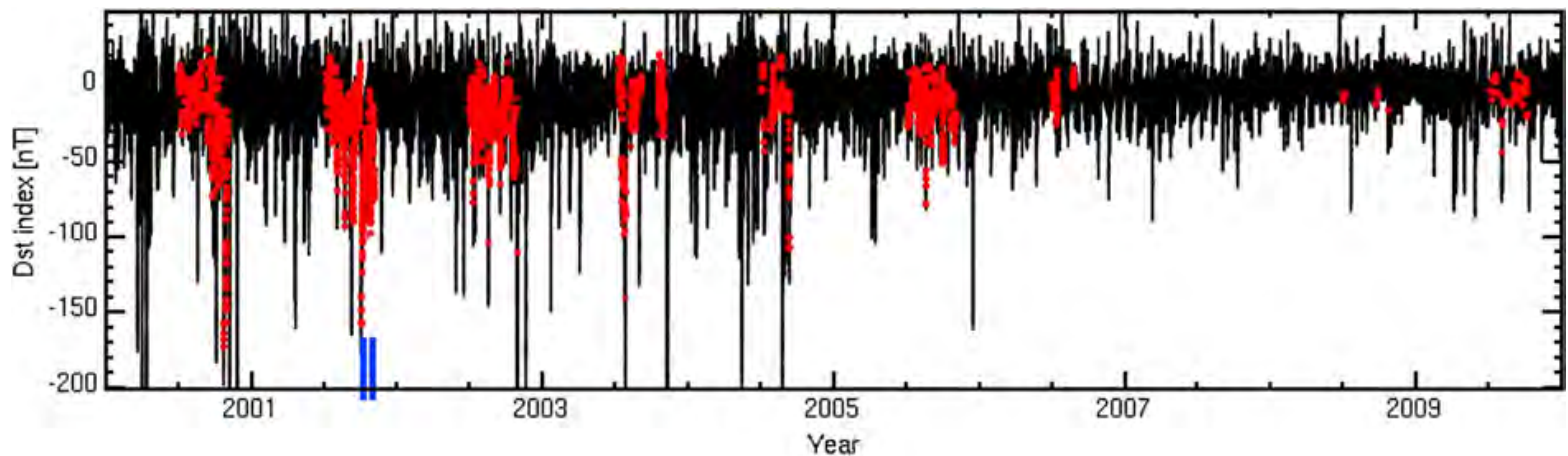

**Figure 2.** Plot of the *SMR* (*Dst*) index for 2001–2010. Red color indicates periods where cold ion observations are available from Cluster. Measurements are only possible when Cluster has apogee in the tail during late July to early November. To be useful for our statistics, we require coverage for at least parts of the main phase and parts of the recovery phase of a storm. These criteria are fulfilled for 32 storms (see Appendix A). The two GEM events in October 2002, discussed in some detail in the present paper, are indicated by blue bars.

The wake data set is somewhat biased toward moderately disturbed conditions. On one hand, utilization of the wake requires a certain bulk velocity (see equation (2)) and minimum solar illumination (see section 2.1). This situation is more likely during slightly disturbed conditions and thus negative *SMR* values. On the other hand, very disturbed conditions with rapidly changing magnetic field will cause the EDI instrument to lose tracking. Consequently, the outflow velocity (equation (5)) cannot be determined. Another reason for less contribution from very disturbed conditions may be that the ions are more often heated to energies above our detection limit. The overall average *SMR* value in our data set is approximately −20 nT, and the average $F_{10.7}$ index is 137. The most intense storm during the years 2001–2010, in terms of *SMR* deflection, took place in October 2003, when the *SMR* index dropped below −500 nT. There are no wake observations during this minimum, so the minimum *SMR* value in our data set is −409 nT, also observed in October 2003.

## 3. Observations of Cold Ion Outflow During Geomagnetic Storms

Geomagnetic storms are large-scale global disturbances in the Earth's magnetic field, typically lasting a couple of days. The time evolution of storms is characterized in terms of three phases referred to as the initial, main, and recovery phases. These stages of storm evolution can be identified from the behavior of the *SMR* (*Dst*) index as illustrated in Figure 3a). For comparison, Figures 3b and 3c show the *SMR* index for the two storms selected by the GEM community for benchmarking and comparison. These will be discussed in some detail in the following subsections.

The initial phase, sometimes referred to as a storm sudden commencement (SSC), is characterized by a positive perturbation in *Dst*, and mainly caused by a compression of the geomagnetic field, often in connection with the arrival of a coronal mass ejection. The compression of the magnetosphere will also cause an increase in the density of the lobes [e.g., *Svenes et al.*, 2008; *Lybekk et al.*, 2012]. Since the cold ion flux is a product of velocity and density (see equation (6)), there will be an apparent increase in outflow during this compression. It is important to note that this apparent increase is primarily a compressional effect, and not necessarily caused by additional supply from the ionosphere.

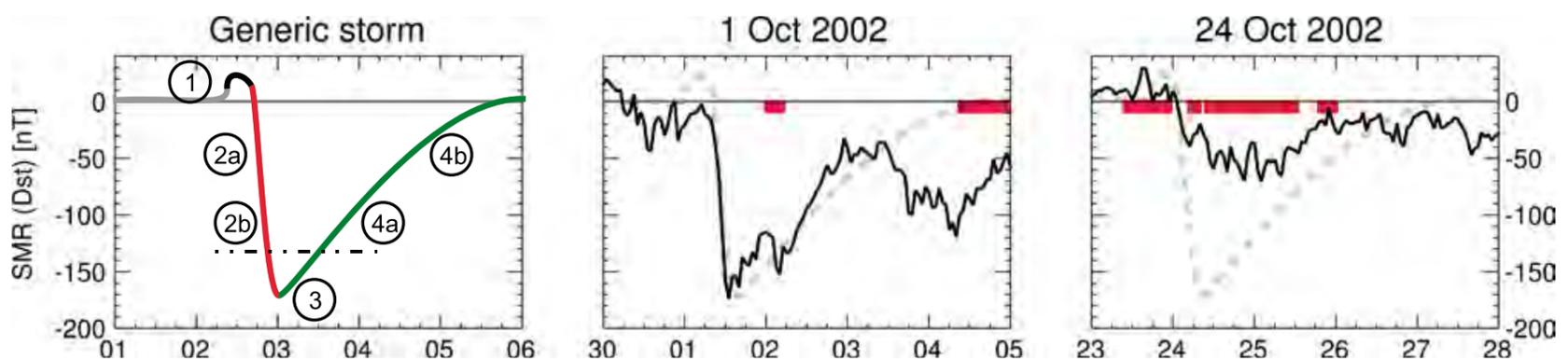

**Figure 3.** (a) Characteristic phases of a geomagnetic storm as manifested in the *SMR* index. The numbered labels indicate stages of evolution and will be used to parametrize a model of the cold ion outflow for a characteristic storm. For comparison, we also show the *SMR* index for each of the geomagnetic storms on (b) 1–5 October 2002 and (c) 23–28 October 2004, respectively, selected by the GEM community for modeling and benchmarking. Red bars indicate intervals where cold ion measurements are available. To guide the reader, dashed gray lines repeat the time profile of the generic storm.

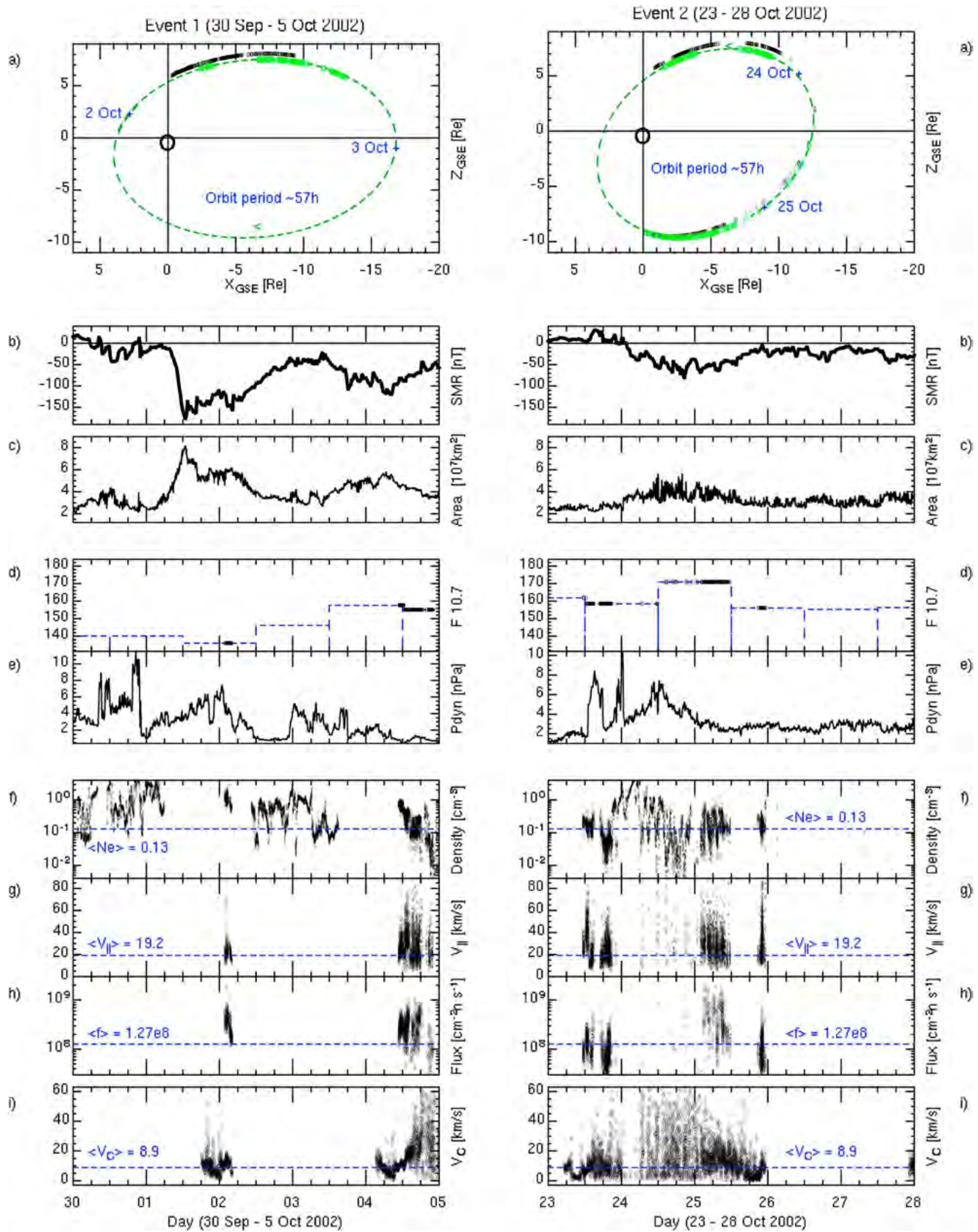

**Figure 4.** Key observations for the two GEM storm intervals discussed in this paper. (left column) Detailed observations for event 1—the geomagnetic storm on 30 September to 5 October 2002. (right column) Same as Figure 4 (left column) but for event 2—the geomagnetic storm on 23–28 October 2002. To facilitate comparison, vertical axis scales are the same as for event 1. (a) $XZ_{GSE}$ projection of the Cluster orbit, (b) *SMR* (*Dst*) index, (c) size of source area, (d) $F_{10.7}$ index, (e) solar wind dynamic pressure, (f) cold ion density, (g) outflow bulk velocity, (h) calculated ionospheric flux, and (i) convection velocity.

The main phase, marked with red color in Figure 3a, is characterized by a significant drop in the *SMR* index over a period of typically 2–10 h. The main phase is a consequence of enhanced transport of plasma inside the magnetosphere and a buildup of energy in the ring current.

During the recovery phase, marked green in Figure 3a, various loss processes will lead to a reduction in the ring current, and *SMR* returns to nonstorm values. Loss processes are slower than the main phase buildup, and a recovery phase can last for several days.

In addition, we introduce an additional stage, labeled ③, which we shall refer to as the "peak phase," which overlaps with the late main and early recovery phases. The peak phase describes the interval where the *SMR* index exceeds 75% of its peak value.

### 3.1. The GEM Storm on 30 September to 5 October 2002

The GEM core group selected two real-world events for model benchmarking and comparisons between models and observations. The first event selected was a storm period starting around 1 October 2002. Figure 4 (left column) gives an overview of observations and some derived quantities during this storm.

As guidelines, horizontal blue dashed lines and values in Figures 4f–4i indicate average (median) quiet time values for density, velocities, and flux. Note that the flux is mapped to ionospheric altitudes (1000 km), so scaling due to flux tube expansion with increasing altitude has been taken into account, but average density and average velocities are based on local measurements at a range of Cluster altitudes.

Wake observations and thus the ability to fully characterize the cold ion outflow during this storm are limited, possibly due to strong heating and thus ion energies above our detection limit. Figure 4a shows the $XZ_{GSE}$ projection of the Cluster C3 orbit (dashed line; Cluster C1 is close nearby) with coverage for C1 indicated as thick black line segments and the coverage C3 as thick green line segments.

The IMF is strongly northward for several hours prior to the storm, and despite a sharp jump in the solar wind dynamic pressure (Figure 4e), the 1–5 October 2002 storm lacks a clear initial phase. As the IMF turns southward around 04 UT on 1 October, a rather large and fast drop (about 180 nT within 8 h) in the *SMR* index, indicating a fast energization of the ring current, is observed. There is only a gradual increase in the solar wind dynamic pressure (Figure 4e) during the main phase, so the storm is primarily driven by strong dayside reconnection following the southward directed IMF. Consequently, an imbalance between dayside reconnection (opening of flux) and nightside reconnection (closing of flux) arises, and the polar cap area, shown in Figure 4c, increases rapidly to almost 3 times its prestorm area.

*SMR* reaches its peak value of −181 nT around noon on 1 October 2002. A second, less pronounced drop in *SMR* is observed around 03 UT on 2 October followed by minor fluctuations in *SMR*. The recovery phase is also interrupted by a new intensification starting on 3 October.

Wake observations are available from two intervals. First, a few hours of observations starting early on 2 October, some hours after the *SMR* minimum of the storm, but still within the stage we have termed peak phase. Cluster is then traversing tailward in the northern lobe. No further wake observations are available until Cluster returns to this region after one orbit (orbit period ≈57 h) on 4 October, corresponding to the recovery of the second intensification. One could argue that the new activation on 3 October should be classified as a new storm. Still, in the text below, we discuss this event as one storm event and refer to the two intervals with observations as the peak phase and the recovery phase of a single storm.

The daily $F_{10.7}$ index, shown as a histogram in Figure 4d, increases from 136 m Wm$^2$ on 2 October to 155 m Wm$^2$ on 4 October 2002. Periods with wake observations are indicated in black and thicker lines. Recall that an increase in $F_{10.7}$ indicates additional solar irradiance and thus potentially more ionization and consequently enhanced cold ion outflow [e.g., *André et al.*, 2015].

Some care must be taken when interpreting the measured density, shown in Figure 4f. Although the highest densities are observed early on 2 October, one should have in mind that these observations are taken closer to Earth (radial distances 6–7 $R_E$) than the later observations on 4 October (radial distances 7–16 $R_E$).

Table 1. Summary of Cold Ion Observations During the First GEM Storm on 1–5 October[a]

| A | B | C | D | E | F | G | H | I | J |
|---|---|---|---|---|---|---|---|---|---|
| | Number | Period With | $\langle Dst \rangle$ | $\langle N_e \rangle$ | $\langle V_\parallel \rangle$ | $\langle$Flux$\rangle$ | PC Area | Outflow Rate | $\langle V_\perp \rangle$ |
| Storm Phase | Records | Data Available | (nT) | (cm$^{-3}$) | (km s$^{-1}$) | ($10^8$ cm$^{-2}$ s$^{-1}$) | ($10^7$ km$^2$) | ($10^{26}$ s$^{-1}$) | (km s$^{-1}$) |
| ③ Peak | 575 | 02 Oct, 01:01-04:04 | −149.0 | 0.88 | 21.0 | 3.06 | 7.44 | 2.22 | 12.9 |
| ④ Recovery | 2310 | 04 Oct, 10:10-22:10 | −95.0 | 0.23 | 27.2 | 2.49 | 5.02 | 1.31 | 17.6 |

[a]Observations were only available during the peak (i.e., SMR (Dst) ≤ −135 nT) and the recovery phases of the storm. The notation ⟨⟩ indicate median values of the respective parameter. The source area in column H is the combined area of Northern and Southern Hemispheres.

The solar wind dynamic pressure is also higher during the first period of observations, and parts of the apparent enhanced density may be due to a compression of the whole magnetosphere as discussed in section 3.

Figure 4g shows outward parallel bulk velocity of the cold ions. Earlier studies using the cold ion data set have shown that the outward parallel velocity increases with geocentric distance [see, e.g., *Engwall et al.*, 2009a, Figure 6]. Indeed, mean and median outflow velocities are lower during the peak phase on 2 October than during the recovery phase on 4 October. This may be related to the findings by *Kitamura et al.* [2013] who found a lower ambient electric field (and thus less acceleration) during main and early recovery phases (which correspond to our peak phase) than during quiet times. However, there are large variabilities, with velocities ranging from 10 to 80 km/s for the peak phase and from 5 to more than 100 km/s during the recovery phase.

Figure 4h shows the mapped flux, i.e., flux at ionospheric altitudes (1000 km) where scaling due to flux tube expansion with increasing altitude has been taken into account. This figure clearly indicates a higher ionospheric outflow flux during the most disturbed period around 2 October than the later observations in the recovery phase on 4 October. Finally, Figure 4i shows 1 min averages of convection. The convection is essentially in the $Z_{GSM}$ direction, i.e., toward the plasma sheet.

Table 1 summarizes the observations. We have here calculated median values over the two stages of the storm where there are observations. The columns are labeled A to J for easy referencing and navigation.

The first period of observation contains a total of 575 records of wake observations and is available during a 3 h period between around 01 UT to 04 UT on 2 October (note that the given time intervals in column C indicate where some data were available, but do not necessarily contain continuous, uninterrupted measurement series, and not necessarily the full time span of the storm phase). The median SMR value for this collection is −149 nT, and this period thus corresponds to the late part of the peak phase, of the storm. This period is characterized by a significantly higher (than nonstorm times) flux and an expanded polar cap region. Consequently, the total outflow rate, $2.22 \cdot 10^{26}$ s$^{-1}$, is also significantly higher than quiet time values (see below) and also significantly higher than values of H$^+$ and O$^+$ outflow reported for disturbed periods in earlier studies [e.g., *Yau et al.*, 1985a].

The second period with wake observations, in total 2310 records over the 10 h period from 10:10 to 22:10 UT on 4 October, is still characterized by a large negative SMR value (median SMR is −95 nT). Both mapped flux and total polar area have decreased since the peak phase, and the resulting outflow is consequently smaller than during the peak phase.

### 3.2. The GEM Storm on 23–28 October 2002

The second storm selected by the GEM community for benchmarking is the result of a corotating interaction region and commences around 15:00 UT on 23 October, with the main phase starting early 24 October 2002. Details are shown in Figure 4 (right column).

Being almost a solar rotation after the first event, the Cluster orbit has precessed about 2 h in local time toward dusk.

This storm is weaker, with a minimum SMR of around −97 nT (Figure 4b). The main phase is longer than for the first event and also shows signatures of individual substorms. There is about 30 h between the SSC on 23 October and the minimum SMR around 20:30 UT on 24 October.

Table 2. As Table 1 but for the GEM Storm on 23–28 Oct 2002

| A | B | C | D | E | F | G | H | I | J |
|---|---|---|---|---|---|---|---|---|---|
| Storm Phase | Number Records | Period With Data Available | $\langle Dst \rangle$ (nT) | $\langle Ne \rangle$ (cm$^{-3}$) | $\langle V_\parallel \rangle$ (km s$^{-1}$) | $\langle$Flux$\rangle$ (10$^8$ cm$^{-2}$s$^{-1}$) | PC Area (10$^7$ km$^2$) | Outflow Rate (10$^{26}$ s$^{-1}$) | $\langle V_\perp \rangle$ (km s$^{-1}$) |
| ① Initial | 1853 | 23 Oct, 11:11-22:10 | −20.0 | 0.08 | 23.2 | 0.89 | 2.97 | 0.35 | 10.1 |
| ② Main | 128 | 24 Oct, 06:06-23:11 | −70.0 | 0.08 | 42.3 | 1.88 | 4.32 | 0.72 | 16.9 |
| ③ Peak | 1523 | 24 Oct, 16:04-25, 09:09 | −84.0 | 0.13 | 28.4 | 2.12 | 4.89 | 0.97 | 17.6 |
| ④ Recovery | 2589 | (24 Oct, 20:08-23:11) | −78.0 | 0.15 | 27.4 | 1.49 | 4.30 | 0.72 | 15.4 |

The variation in source area (Figure 4c) is also much smaller in this case. From an initial size of just above 10$^7$ km$^2$ prior to the storm, the total polar cap area expands to about 4·10$^7$ km$^2$ around the peak phase on 24 October.

The solar wind dynamic pressure shows a very similar behavior as the previous event, with an initial pressure pulse and a gradual increase during the first half of the main phase. $F_{10.7}$, and thus ionization, is highest during the peak phase of the storm (though there is probably no direct causal relation between the $F_{10.7}$ index and storm phase).

Observations of cold ion outflow, shown in Figures 4g and 4h, are available from around 10:27 UT on 23 Oct when Cluster was in the Southern Hemisphere until around 23:55 UT on 25 October. All measurements were taken between 6 and 19 $R_E$ geocentric distance, and unlike the previous events, we have observations from all phases of the storm for this event. As for the previous event, there is significant spread in the measurements. Perhaps, the most pronounced feature in the observations is the distinctly higher flux (4h) during the peak phase.

We also note that the plasma convection (Figure 4i) picks up rapidly as the main phase of the storm progresses and subsides as the storm abates.

Table 2 lists averages for this storm. Despite similar solar wind dynamic pressure values, average cold ion densities are consistently lower than for event 1. With exception of the main phase, from which there are only 128 records with wake observations, outflow velocities are fairly constant and in the same range as for event 1.

### 3.3. A Generic Geomagnetic Storm

A typical geomagnetic storm lasts a couple of days. Wake characterization, and thus cold ion outflow measurements are only available for at best a few hours when the Cluster satellites traverses the lobes, and often only intermittently. The two selected GEM events above are examples of this. Thus, we do not have full coverage of cold ion observations throughout any of the storms in Figure 2. Still, we can combine observations from several storms to gain knowledge about cold ion outflow during storms in general.

During the years 2001–2010, we visually identified a total 32 geomagnetic storms where cold ion data were available for at least some intervals in both the main and recovery phases. For each of these storms, we recorded start times and durations of the various storm phases and added this information to the cold ion database. The peak (minimum) value of the SMR index for each storm was also noted.

For better parametrization, we made a further division of the storm evolution. In Figure 3a we have labeled these stages with numbers ① to ④. Not all storms exhibit a pronounced initial phase (labeled ①), so we will not focus much on this stage.

First, we divide the main phase into an early and a late stage (labeled ②a and ②b), where the early stage contains the first half of the SMR drop until minimum, and the second stage is the time until the peak SMR is reached.

We do a similar division of the recovery phase. The label ④a refers to the early recovery, and the label ④b refers to the late recovery stage. The classification of this latter stage is subject to some uncertainty, as it is not always easy to accurately determine when the effects of a storm have fully subsided. There are also intervals where a new storm commences in what appears to be the late recovery of an earlier storm.

As noted above, we have introduced an additional stage (the peak phase) labeled ③, which describes the interval where the SMR index exceeds 75% of its peak value. The peak phase consists of the late main and early recovery phases.

**Table 3.** Similar to Tables 1 and 2 but Now With Characteristics for the Full Data Set[a]

| A | B | C | D | E | F | G | H | I | J |
|---|---|---|---|---|---|---|---|---|---|
| Storm Phase | Number Records | ⟨Duration⟩ (h) | ⟨Dst⟩ (nT) | ⟨Ne⟩ (cm$^{-3}$) | ⟨$V_{\parallel}$⟩ (km s$^{-1}$) | ⟨Flux⟩ (10$^8$ cm$^{-2}$ s$^{-1}$) | PC Area (10$^7$ km$^2$) | Outflow Rate (10$^{26}$ s$^{-1}$) | ⟨$V_{\perp}$⟩ (km s$^{-1}$) |
| Nonstorm | 10,824 | - | 6.0 | 0.13 | 19.2 | 1.27 | 2.61 | 0.33 | 8.9 |
| ① Initial/SSC | 10,356 | 4.0 | −7.0 | 0.13 | 22.7 | 3.35 | 2.98 | 0.52 | 11.1 |
| ② Main | 18,771 | 5.0 | −39.0 | 0.34 | 29.2 | 3.35 | 4.38 | 1.66 | 14.3 |
| ⓐ - Early | 7,801 | - | −17.0 | 0.23 | 27.9 | 2.26 | 3.81 | 0.88 | 14.3 |
| ⓑ - Late | 10,970 | - | −48.0 | 0.43 | 30.1 | 4.58 | 4.65 | 3.08 | 14.3 |
| ③ Peak | 21,535 | 12.0 | −65.0 | 0.42 | 36.7 | 5.31 | 4.50 | 2.73 | 16.3 |
| ④ Recovery | 110,815 | 84.0 | −34.0 | 0.11 | 25.3 | 1.58 | 3.56 | 0.58 | 12.9 |
| ⓐ - Early | 52,748 | - | −60.0 | 0.15 | 27.6 | 2.33 | 3.92 | 0.98 | 15.1 |
| ⓑ - Late | 58,067 | - | −18.0 | 0.08 | 23.0 | 1.19 | 3.10 | 0.36 | 10.7 |

[a]Rather than individual times, we now provide the average duration of the various stages of storm evolution in column C. Not all storms had a pronounced initial phase or storm sudden commencement, so estimates for this phase are less reliable. This table defines our "generic storm."

Our generic storm is simply constructed by the taking the averages (medians) of the various parameters over these 32 storms. Table 3 summarizes the data characteristics of our generic storm. For comparison, we also include quiet time periods, where "quiet" is simply defined as all intervals where the *SMR* index is positive, and no storms were identified.

We also investigated whether there are any fundamental differences in ion outflow between weak and strong storms (several *SMR* ranges were tested to define "weak" and "strong"). Densities, outflow velocities, and fluxes were lower for weak storms, but the general behavior of increasing outflow as the main phase progresses, strongest outflow during the peak phase and a gradual abatement during the recovery phase did not change.

## 4. Discussion

Figure 4 together with Tables 1–3 draw a fairly consistent picture of cold ion outflow during geomagnetic storms: Cold ion outflow increases with increasing storm intensity, and the largest outflow is observed around the peak phase of a storm. During the recovery phase, the outflow subsides but is still stronger than during quiet times. Below, we discuss the observations, in particular, those listed in Table 3, in some detail, and try to identify processes responsible for the observed bahavior.

### 4.1. Variations in Density and Velocity During Storms

The observations above indicate that the density (column E in Tables 1–3) increases throughout the main phase, reaches a maximum when the storm reaches its late main phase or peak phase, and decreases as the effects of the storm subside during the recovery phase.

As mentioned above, some caution is required when interpreting density values. In addition to the altitude of the observations, variations in the observed density can be due to several processes, of which the most probable are (1) a genuine increase in the supply of plasma from the ionosphere, (2) compression of the whole magnetosphere, and (3) "contamination" through inflow of magnetosheath plasma following dayside or high-latitude reconnection.

Ideally, our measurements should only be sensitive to the first process, but there is no way to actually identify the relative contributions of the above processes. Magnetosheath plasma typically has higher temperatures, but recall that the cold plasma density is derived from the spacecraft voltage, and even magnetosheath-like plasma can affect the spacecraft voltage. Contamination from other sources can therefore not be excluded [e.g., *Pedersen et al.*, 2001].

There is only a weak correlation between the solar wind dynamic pressure and the observed density (linear Pearson correlation coefficient = 0.26), so compression of the magnetosphere alone cannot explain the observed variations in cold ion density throughout storms seen in Table 3.

Increased Joule heating in the ionosphere [e.g., *Rodger et al.*, 2001] and other heating processes raise the scale height of both neutral and ionized components of the thermosphere, resulting in increased ion density above the exobase and thus increase the source reservoir. This may explain some of the observed density variations throughout storms. During very disturbed conditions, a larger fraction of the ions will probably be heated to energies above our detection limit ($\approx$70 eV) though.

Variations in bulk outflow velocity (column F) are smaller than the density variations. Table 3 shows that the bulk velocity is highest during the peak phase of the storm. Parallel bulk flow velocities are derived from wake observations, and thus only sensitive to cold ions (see section 2.1.2). They are thus less ambiguous than density values. We note that the maximum velocities in Tables 1 and 2 seem to be in the recovery and main phases, respectively. The limited data coverage may be one possible reason for this apparent discrepancy.

At low altitudes, acceleration is primarily caused by an imbalance between the downward gravity and upward forces from the ambient electric fields and the mirror force. Probably, neither of these forces vary greatly throughout storm phases, however. From the results of *Kitamura et al.* [2013], one would even expect the ambient electric field in the polar cap region to be smaller during disturbed periods. During periods with high geomagnetic activity, additional parallel fields may play a role near the cusp and auroral zone, but less so in the open polar cap area.

At Cluster altitudes, centrifugal acceleration, although small, is the only relevant force. Centrifugal acceleration is governed by magnetospheric convection which is higher during disturbed periods. We thus argue that the observed variations in velocity throughout the storm phases are a result of field-aligned acceleration caused by centrifugal forces working over long distances. This is corroborated by the fact that the parallel velocity is more or less proportional to the convection in the tables.

Note that the low cold ion outflow velocities imply long transport times (typically of the order of hours) from the ionosphere to the lobes where our observations are made. Although these transport times are smaller than typical timescales of the various storm phases, it might nonetheless be important for modelers.

### 4.2. Variations in Source Area and Flux of Cold Ions

The mapped flux (column G) is a product of the locally measured density and bulk velocity but mapped to ionospheric heights (1000 km). The mapped flux show strong variations as the storm progresses, with the maximum flux, $5.31 \cdot 10^8$ cm$^{-2}$ s$^{-1}$, at the peak phase of the storm. This value is more than 3 times higher than the quiet time flux. As the storm subsides, the flux decreases.

The source area given in column H in Tables 1–3 is calculated using the assumption that the open polar cap is the source of the cold ions. We see the same trend in the polar cap area as the other parameters; the source area increases in size as the intensity of the storm increases and reaches a maximum when the storm is at its most intense and decreases as the storm subsides.

Models [e.g., *Cully et al.*, 2003; *Ebihara et al.*, 2006] and observations [e.g., *Moore et al.*, 1999; *Haaland et al.*, 2012a; *Li et al.*, 2013] have shown that the fate of outflowing ions is largely governed by the convection. High convection means a faster transport to the plasma sheet and essentially no ion escape direct into the solar wind along open field lines.

The observations summarized in Tables 1–3 are taken at a range of altitudes. Due to the evolution of Cluster's orbit, with the line of apsides moving down as time progresses, there are more observations from Southern Hemisphere. Southern Hemisphere observations are on average taken at higher altitudes than Northern Hemisphere observations.

Since we scale flux values to ionospheric altitudes, the actual fluxes are not affected, but there will be an orbital bias: Consider the situation illustrated schematically in Figure 5. During quiet periods with low or stagnant convection, the effective transport path for cold ions from a given dayside source will be along the blue arrow. This outflow will be detected by a spacecraft located in the vicinity of region B, i.e., rather high altitudes and over an extended time period.

During disturbed condition, the convection is stronger, and the effective transport path will be along the orange arrow and can only be detected when Cluster is around region A, i.e., at lower altitudes and for shorter time periods.

This bias is also present in our characteristic data set in Table 3. Main phase and peak phase measurements are on average taken at lower altitudes than during quiet times and recovery time observations.

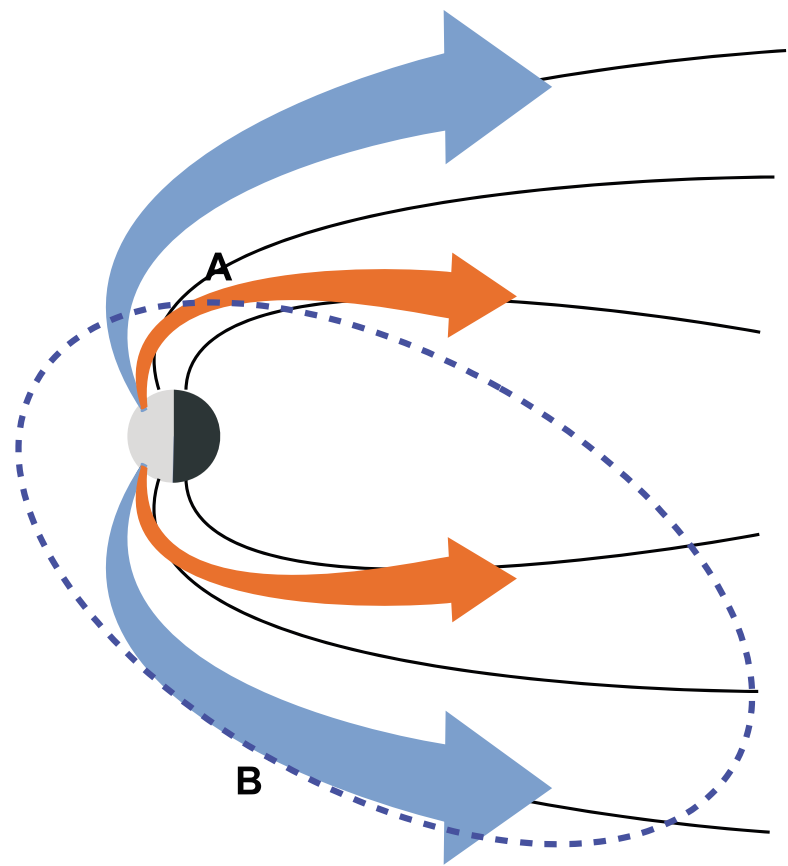

**Figure 5.** Illustration of orbital bias. Recall that wake measurements are only available in the high-latitude, nightside lobe regions. Due to the evolution of Cluster's orbit (blue dashed line), with apogee moving farther into the southern lobe as the years pass, we are more likely to observe outflow from a given source location in the ionosphere during quiet periods (blue transport path) in the Southern Hemisphere and in the Northern Hemisphere during disturbed periods (fast convection, orange outflow transport path).

### 4.3. Accumulated Outflow During a Storm

Table 3 also allows us to estimate the total cold ion outflow throughout a storm. Our generic storm has a duration of almost 100 h (column C in Table 3, but taking into account that the peak phase overlaps with the main and recovery phases). If we take the durations of the individual phases and multiply with the respective average outflow rates (column I), we obtain a total outflow of approximately $7 \cdot 10^{32}$ ions. For comparison, the nonstorm total outflow over the same time period would be of the order $3.5 \cdot 10^{32}$ ions.

Since storm times are associated with enhanced convection (column J), the outflow is more likely to be transported to the plasma sheet [*Haaland et al.*, 2012a; *Li et al.*, 2013], whereas during stagnant convection, a larger fraction of the ouflowing ions are lost downtail into the solar wind. The technique used in present study is not able to resolve composition, but earlier results [see e.g., *Kistler et al.*, 2006, and references therein] indicate that the $O^+$ abundance and thus the $O^+/H^+$ ratio increase significantly during storm times. In terms of mass transport, the supply to the near-Earth plasma sheet is therefore much larger than the factor 2 change in cold ion outflow between quiet time and storm time.

## 5. Summary

We have presented observations of cold ion outflow during two selected geomagnetic storm events and calculated characteristic outflow parameters which may be useful for benchmarking against models and simulations. The observational results can be summarized as follows:

1. At a given location, cold ion density in the lobe region varies with storm intensity. Higher geomagnetic activity (characterized by larger negative *SMR* values) is associated with higher cold ion densities. Average lobe densities at Cluster altitudes (4–19 $R_E$) vary between 0.13 cm$^{-3}$ during quiet times to about 0.4 cm$^{-3}$ during disturbed periods.
2. Variations in bulk outflow velocity also show correlation with storm intensity, although the variations are typically 50% or less between the lowest outflow velocities observed during quiet times and the highest outflow velocities observed during the peak intensity of the storm. The increased bulk outflow velocity is probably a result of larger centrifugal forces due to enhanced convection during disturbed conditions.

| Table A1. List of Storms Used to Generate Our Characteristic Storm Described in Section 3.3 and Table 3 | | |
|---|---|---|
| Year | Storm Interval | $SMR_{min}$ |
| 2001 | 19 Aug 15:07–21 Aug 11:07 | −133 |
| 2001 | 13 Sep 13:21–14 Sep 02:21 | −58 |
| 2001 | 23 Sep 13:51–25 Sep 16:51 | −83 |
| 2001 | 28 Sep 07:13–30 Sep 13:13 | −116 |
| 2001 | 3 Oct 02:25–4 Oct 19:25 | −122 |
| 2001 | 5 Oct 11:36–7 Oct 07:36 | −184 |
| 2001 | 11 Oct 21:02–12 Oct 00:02 | −73 |
| 2001 | 12 Oct 05:38–14 Oct 07:38 | −79 |
| 2001 | 24 Oct 12:39–26 Oct 19:39 | −211 |
| 2001 | 28 Oct 03:43–30 Oct 19:43 | −140 |
| 2002 | 5 Jul 17:36–8 Jul 10:36 | −54 |
| 2002 | 12 Jul 21:24–14 Jul 20:24 | −48 |
| 2002 | 1 Aug 10:11–1 Aug 18:11 | −52 |
| 2002 | 2 Aug 23:11–3 Aug 20:11 | −78 |
| 2002 | 4 Aug 03:46–5 Aug 02:46 | −49 |
| 2002 | 19 Aug 23:06–23 Aug 23:06 | −96 |
| 2002 | 10 Sep 16:18–18 Sep 07:18 | −166 |
| 2002 | 4 Oct 10:45–13 Oct 15:45 | −181 |
| 2002 | 14 Oct 05:09–15 Oct 11:09 | −94 |
| 2002 | 15 Oct 16:30–16 Oct 14:30 | −46 |
| 2002 | 24 Oct 06:31–29 Oct 02:31 | −85 |
| 2002 | 28 Oct 05:01–30 Oct 03:01 | −48 |
| 2003 | 12 Jul 15:04–14 Jul 15:04 | −108 |
| 2003 | 17 Jul 03:28–19 Jul 09:28 | −106 |
| 2003 | 26 Jul 18:02–27 Jul 17:02 | −62 |
| 2003 | 6 Aug 12:19–7 Aug 16:19 | −63 |
| 2003 | 19 Aug 09:20–21 Aug 06:20 | −148 |
| 2003 | 21 Aug 04:43–25 Aug 10:43 | −59 |
| 2003 | 24 Sep 00:40–27 Sep 14:40 | −48 |
| 2003 | 2 Oct 16:15–4 Oct 20:15 | −52 |
| 2003 | 17 Oct 18:58–20 Oct 14:58 | −103 |
| 2003 | 30 Oct 19:23–3 Nov 00:23 | −409 |
| 2004 | 16 Jul 22:53–19 Jul 10:53 | −96 |
| 2004 | 22 Jul 21:19–23 Jul 18:19 | −104 |
| 2004 | 25 Jul 22:45–26 Jul 23:45 | −149 |
| 2004 | 28 Jul 02:23–31 Jul 04:23 | −234 |
| 2004 | 3 Sep 20:13–7 Sep 09:13 | −133 |
| 2004 | 20 Oct 03:46–22 Oct 15:46 | −47 |
| 2005 | 9 Jul 10:52–10 Jul 02:52 | −57 |
| 2005 | 29 Jul 23:03–1 Aug 05:03 | −45 |
| 2005 | 11 Sep 13:32–15 Sep 08:32 | −132 |
| 2006 | 5 Jul 16:13–9 Jul 22:13 | −50 |
| 2006 | 29 Jul 12:09–31 Jul 17:09 | −59 |
| 2006 | 7 Aug 23:38–11 Aug 01:38 | −53 |
| 2006 | 19 Aug 13:21–24 Aug 02:21 | −83 |
| 2006 | 5 Sep 19:31–10 Sep 00:31 | −56 |

**Table A1.** (continued)

| Year | Storm Interval | $SMR_{min}$ |
|---|---|---|
| 2006 | 30 Sep 18:05 – 4 Oct 09:05 | −49 |
| 2007 | 14 Jul 10:40 – 17 Jul 14:40 | −48 |
| 2009 | 23 Oct 17:41 – 25 Oct 00:41 | −46 |
| 2010 | 6 Aug 11:21 – 9 Aug 12:21 | −79 |

3. The flux of cold ions from the ionosphere is of the order of $1 \cdot 10^8$ cm$^{-2}$ s$^{-1}$ during quiet times and more than 5 times higher during storm maximum.
4. The source area, assumed to be the open polar cap regions, varies significantly with storm intensity. During the peak phase of the storm, the source area is typically almost twice as large as the quiet time area.
5. Outflow rates vary almost an order of magnitude between quiet and very disturbed condition. The average quiet time outflow was $0.3 \cdot 10^{26}$ s$^{-1}$, increasing to a maximum of $2.7 \cdot 10^{26}$ s$^{-1}$, during the peak phase of our characteristic storm.
6. During a typical storm (i.e., our "generic storm"), the total accumulated outflow is of the order $7 \cdot 10^{32}$ ions, which is roughly twice as much as during quiet time conditions.
7. During disturbed periods, convection is stronger, and the outflowing cold ions are more likely to be supplied to the near-Earth plasma tail.

## Appendix A: Storm List

Table A1 lists the dates and times of geomagnetic storm periods used to estimate characteristic cold ion outflow key parameters during geomagnetic storms. Note that this list only shows intervals of storms where we have some observations during the main and recovery phases. In general, we do not have full coverage during a storm. Cold ion data are only available during late July to early November when Cluster has its apogee in the geomagnetic tail.

**Acknowledgments**

This study was supported by the Norwegian Academy of Science and Letters and the Fritjof Nansen Fund for Scientific Research. We also thank the staff at EISCAT Svalbard for their support during field work in Svalbard. Computer code used for the calculations in this paper has been made available as part of the QSAS science analysis system. Solar wind data were obtained from the Coordinated Data Analysis Web (CDAWeb—see http://cdaweb.gsfc.nasa.gov/about.html). We also thank the International Space Science Institute, Bern, Switzerland, for providing computer resources and infrastructure for data exchange.